\documentclass[aps,showpacs,twocolumn,nofootinbib]{revtex4}
\usepackage{amssymb}
\usepackage{natbib}
\usepackage{amsmath}
\usepackage{amsfonts}
\usepackage{graphicx}
\usepackage{mathrsfs}
\usepackage{dcolumn}
\usepackage{bm}
\usepackage{color}
\definecolor{rot}{rgb}{0.75,0.05,0.25}
\definecolor{hellgrau}{gray}{0.5}
\definecolor{blau}{rgb}{0,0,0.7}

\def\Tr{\mbox{Tr}}

\begin{document}

\title{Notes on heat engines and negative temperatures}

\author{Michele Campisi}
\email{michele.campisi@sns.it}
\affiliation{NEST, Scuola Normale Superiore \& Istituto Nanoscienze-CNR, I-56126 Pisa, Italy}
\date{\today }
\begin{abstract}
We show that a Carnot cycle operating between a positive canonical-temperature bath and a negative canonical-temperature bath has efficiency equal to unity. It follows that a negative canonical-temperature cannot be identified with an absolute temperature. We illustrate this with a spin in a varying magnetic field.
\end{abstract}
\pacs{
} 
 \maketitle
\subsection{A general remark on the impossibility of efficiency larger than one}
In the context of the issue of negative temperatures it is often claimed that
negative temperatures imply Carnot efficiencies larger than one \cite{Braun13SCIENCE339,Frenkel15AJP83}. Such claims are based on the use of the Carnot engine formula for the efficiency
\begin{align}
\eta = 1-T_C/T_H
\label{eq:etaC}
\end{align}
with a positive $T_C$ and a negative $T_H$. This procedure is erroneous for a simple reason:  In deriving Eq. (\ref{eq:etaC}) one uses the assumption that $T_C$ and $T_H$ have the same sign. Hence the formula cannot be used with two temperatures of opposite signs. We recall the derivation of Eq. (\ref{eq:etaC}) for convenience. I should emphasise here that this is a basic thermodynamic derivation that does not refer to any specific statistical ensemble. Consider a Carnot cycle. Consider first the isothermal expansion at $T_H$. The heat entering the system is $Q_H= T_H \Delta S_H$. The adiabatic expansion has $ Q= 0$, hence $\Delta S=0$. With the isothermal compression, the system must go back to the very initial entropy, hence $\Delta S_C = -\Delta S_H = Q_C/T_C$. Therefore $Q_C = - Q_H T_C/T_H$. Imagine $Q_H>0$. If $T_C$ and $T_H$ have equal sign, then $Q_C<0$. So the heat intake $Q_\text{in}$ is given by $Q_\text{in}=Q_H$. Using the first law of thermodynamics (total work output $W$ is given by the total heat balance $Q_\text{in}+Q_\text{out}=W$) and the efficiency definition
\begin{align}
\eta = W/Q_\text{in}
\end{align}
Eq. (\ref{eq:etaC}) immediately follows. Imagine now instead that $T_C$ and $T_H$ have opposite signs, then heat enters the system in both isotherms \cite{Ramsey56PR103}, hence $Q_\text{in}=Q_H+Q_C=W$, and the correct formula is now 
\begin{align}
\eta =1 \quad [\text{sign}(T_C) \neq \text{sign}(T_H)]
\end{align}
Mis-identification of $\eta$ with $W/Q_H$, as in Ref. \cite{Braun13SCIENCE339,Frenkel15AJP83} would lead to the absurd conclusion that $\eta>1$.

The very concept on an absolute scale of temperatures is constructed upon Eq. (\ref{eq:etaC}) \cite{Uffink01SHPMP32}. Since Eq. (\ref{eq:etaC}) only holds under the provision that  both quantities $T_H, T_C$ have same sign (conventionally positive), there apparently is no room for interpreting any negative quantity, e.g. a negative canonical-temperature (see below), as an absolute temperature. 

\subsection{A canonical Carnot cycle involving a negative canonical temperature}

We consider a canonical Carnot cycle where two isothermal transformations are alternated by two adiabatic transformations. By ``canonical cycle'', we mean that the system is in a canonical Gibbs state at all times during the cycle.
The isothermal transformations take place while the system stays in contact with a thermal bath characterised by the given inverse canonical-temperature (c-temperature) $\beta$. A thermal bath at inverse c-temperature $\beta$ is a physical system with the property of leading a system of interest to the state 
$e^{-\beta H(\lambda)}/Z(\lambda,\beta)$ when the two are allowed to interact for a sufficiently long time.
$ H(\lambda)$ is the Hamiltonian of the system of interest, which depends on the work-parameter $\lambda$.
We leave aside the question if a thermal bath with negative $\beta$ can exist or can be engineered.

\begin{figure}[t]
		\includegraphics[width=.4\textwidth]{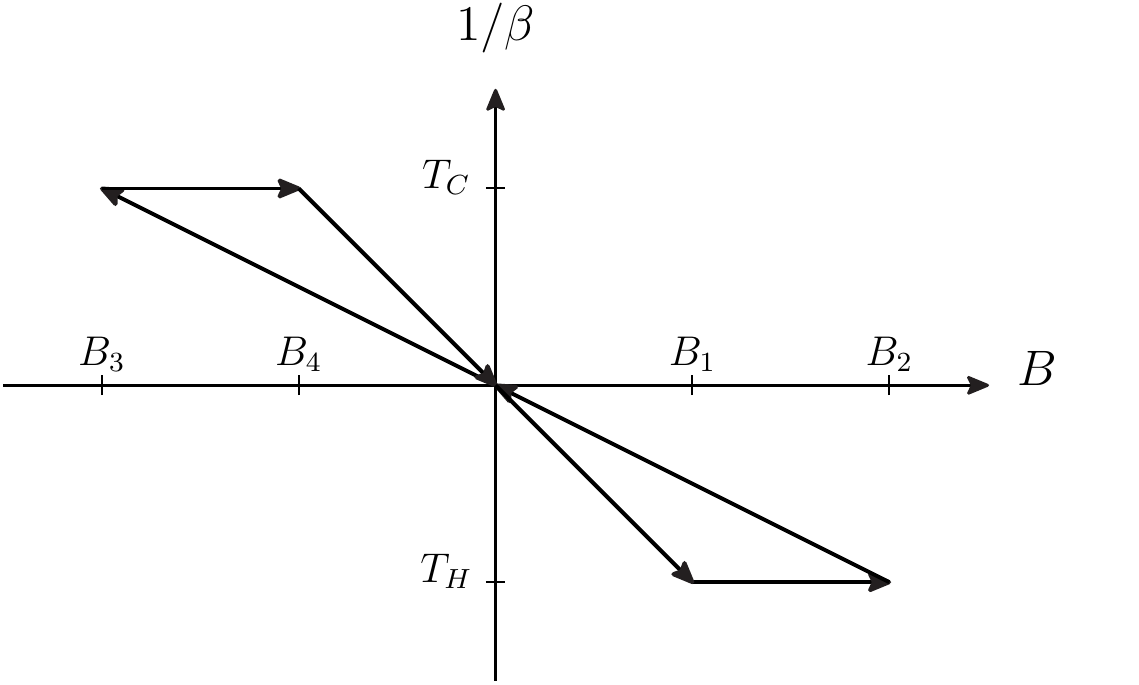}
		\caption{A canonical Carnot cycle of a spin $1/2$ between baths of opposite canonical temperature sign. The cycle can be decomposed in two cycles each of unit efficiency. The overall efficiency is therefore $\eta=1$.}		\label{fig:Fig3}
\end{figure}

A Carnot cycle between opposite c-temperature baths is illustrated in 
Fig. \ref{fig:Fig3} for the case of a single spin $1/2$ with Hamiltonian $H(B) = -B \sigma^z/2$, with $\sigma_z$ a Pauli operator. The magnetic field $B$ plays the role of work parameter. The system is prepared at c-temperature $T_H<0$ and $B_1>0$, its state is $\rho_1\propto e^{-B_1 \sigma^z/k_B T_H}$. It undergoes then an isothermal transformation of the magnetic field to $B_2>B_1$ at c-temperature $T_H$, thus reaching the state $\rho_2\propto e^{-B_2 \sigma^z/k_B T_H}$. Contact with the thermal reservoir is now removed and the magnetic field is brought to $B_3<0$. Independent of the speed of the magnetic field reversal, no jumps between the two spin states occur because the spin Hamiltonian commutes with itself at all times. Accordingly the state $\rho$ remains unvaried and the new positive c-temperature $T_C=T_H B_3/B_2>0$ is reached:
$\rho_3=\rho_2=\propto e^{-B_3 \sigma^z/k_B T_C}$. This reversal does not suffer the problems of passage through null c-temperature mentioned in  \cite{Ramsey56PR103,Dunkel14NATPHYS10,Frenkel15AJP83}.
The spin is now brought into contact with the bath at c-temperature $T_C$ and the magnetic field is isothermally switched to $B_4=B_1 T_C/T_H<0$, so that the state $\rho_4\propto e^{-B_4 \sigma^z/k_B T_H} = e^{-B_1 \sigma^z/k_B T_C}=\rho_1$ is reached. A switch of the magnetic field back to $B_1$ after thermal contact is removed, closes the Carnot cycle. Using the quantum mechanical formula $U = \Tr \rho H$ it is straightforward to calculate the internal energy of each state and accordingly the heat exchanged with each bath. One finds, at variance with the ordinary Carnot engine which withdraws energy from the hot bath only, that this engine withdraws heat from both baths $Q_H>0, Q_C>0$. Hence the heat intake is $Q_\text{in}= Q_H+ Q_C=W$ and the efficiency is one:
$
\eta = 1
$.

This result can also be obtained by doing no math by noting that the Cycle is composed of two sub-Carnot-cycles, one operating between c-temperatures $0$ and $T_C$ (with efficiency $\eta=1-0/T_C =1$), and the other between $T_H$ and $0$ (also with efficiency $1- 0/T_H=1$ \cite{Ramsey56PR103}). The overall efficiency is therefore $1$.

\subsection*{Acknowledgements}
This research was supported by a Marie Curie Intra European Fellowship within the 7th European Community Framework Programme through the project NeQuFlux grant n. 623085 and by the COST action MP1209 ``Thermodynamics in the quantum regime''.

\end{document}